\documentclass[aps,preprint,superscriptaddress,nofootinbib]{revtex4-1}
\usepackage{graphicx,amssymb,amsmath,hyperref,cancel,xcolor,bm,multirow}

\newcommand{\la}{\lambda}
\newcommand{\be}{\beta}
\renewcommand{\ol}{\overline}
\newcommand{\wt}{\widetilde}
\newcommand{\nl}{\nonumber\\}
\newcommand{\lt}{\left}
\newcommand{\rt}{\right}
\newcommand{\eq}[1]{\begin{align}#1\end{align}}

\begin{document}

\title{Dirac neutrino from the breaking of Peccei-Quinn symmetry}
\author{Seungwon Baek}
\affiliation{Department of Physics, Korea University, Seoul 02841, Korea}

\begin{abstract}
We propose a model where Dirac neutrino mass is obtained from small vacuum expectation value (VEV) of neutrino-specific Higgs doublet without fine-tuning problem. The small VEV results from a seesaw-like formula with the high energy scale identified as the Peccei-Quinn (PQ) symmetry breaking scale. Axion can be introduced {\it \`a la} KSVZ or DFSZ. The model suggests  neutrino mass, solution to the strong CP problem, and dark matter may be mutually interconnected.
\end{abstract}
\maketitle

%%%%%%%%%%%%%%%%%%%%%%%%%%%%%%%%%%%%%%%%%%%%%%%%%%%%%%%%%%%%%%%%%%%%%%
\section{Introduction}\label{sec:introduction}

Neutrino mass, strong CP problem, and the existence of dark matter are some hints that call for new physics (NP) beyond the
standard model (SM). In this paper we consider a new physics (NP) model which can address these three problems simultaneously
without fine-tuning.

Neutrino mass can be generated in neutrino-specific two Higgs doublet model ($\nu$THDM) where one Higgs doublet $\Phi_1$ with VEV 
$v_1(\sim 1 \,{\rm eV})$ couples only to lepton doublet and right-handed neutrinos
 and the other Higgs doublet $\Phi_2$ with VEV $v_2(=246\, {\rm GeV})$ couples to all the other quarks and 
 charged-leptons~\cite{Ma:2000cc,Grimus:2009mm,Wang:2006jy,Davidson:2009ha,Ma:2016mwh,Baek:2016wml}. We assume a global $U(1)_X$ symmetry under which
 $\nu_R$ and $\Phi_1$ are charged. The symmetry prohibits the mass term for the right-handed neutrinos. 
 Therefore the neutrino gets only Dirac mass term and its Yukawa coupling can be of order one.
 The tiny VEV $v_1$ necessary to explain neutrino mass can be generated by seesaw-like relation
 in which the high-energy scale is the electroweak scale~\cite{Davidson:2009ha,Ma:2016mwh,Baek:2016wml}. 
The scalar $S$ which is a SM-singlet but charged under $U(1)_X$ breaks the global symmetry spontaneously,
and can couple to an electroweak-scale WIMP dark matter which is stabilized by a  remnant discrete symmetry~\cite{Ma:2016mwh,Baek:2016wml,Baek:2018wuo}. 
 
 In this paper we consider a scenario in which the VEV $v_S$ of  $S$ and the mass scale of $\Phi_1$ is lifted to
 a very large scale $\sim 10^{12}$ GeV. 
 The neutrino mass is generated by a mechanism shown in Figure~\ref{fig:neutrino_mass}.
 The diagram generates a VEV $v_1$, which can be written as
 \begin{align}
 v_1 \approx {\sqrt{2} \mu v_2  v_S \over m_{11}^2},
 \end{align}
 where $v_2=246$ GeV, $v_S \sim m_{11} \sim {\cal O}(10^{12})$ GeV, $m_{11}$ being the mass scale of $\Phi_1$.
 We extend the model to incorporate axions so as to solve the strong CP problem and dark matter candidate.
  In this case the $U(1)_X$ is identified with the Peccei-Quinn (PQ) symmetry $U(1)_{\rm PQ}$, 
  and after $S$ getting VEV the Nambu-Goldstone boson becomes an axion.
  Therefore the neutrino mass and the axion are connected by the scalar $S$.
  Since $m_{11} \sim {\cal O}(10^{12})$ GeV, the low energy constraints such as collider searches and charged lepton number violating processes are irrelevant.
Since  the axion is also a good cold dark matter candidate, the scenario also solves the dark matter problem with the axion as a cold dark matter.
Symmetry arguments show that the hierarchy $v_1 (\sim {\cal O}(1\, {\rm eV})) \ll v_2(\simeq246\, {\rm GeV}) \ll v_S (\sim {\cal O}(10^{12}\, {\rm GeV}))$ 
is technically natural.
%For our purpose we take $v_1 \sim {\cal O}(10\, {\rm eV})$, $v_2 = 246$ GeV, and $v_S \sim {\cal O}(10^{12}\, {\rm GeV})$.

The $\nu$THDM on its own does not provide an axion candidate. We make the Nambu-Goldstone boson coming from the spontaneous
breaking of the $U(1)_X$ in the model of \cite{Baek:2016wml}
an axion by introducing either heavy vector-like quarks $\Psi_{L,R}$ (KSVZ-type axion) or additional Higgs doublet (DFSZ-type axion).
It turns out that the phenomenology of the axion in the model is very close to that of original KSVZ~\cite{Kim:1979if,Shifman:1979if}
 and DFSZ~\cite{Dine:1981rt,Zhitnitsky:1980tq} axion models, respectively.
 There are many attempts to connect axion and neutrino mass in the literature~\cite{
 Berezhiani:1989fp,
 Gu:2006dc,
 Chen:2012baa,
 Dasgupta:2013cwa,
 Ma:2014yka,
 Bertolini:2014aia,
 Ahn:2015pia,
 Gu:2016hxh,
 Ma:2017zyb,
Suematsu:2017kcu,
Ahn:2018cau,Reig:2018yfd,Carvajal:2018ohk}.

The paper is organized as follows.
In Section~\ref{sec:model}, we briefly review the $U(1)_X$ model studied in~\cite{Baek:2016wml} with $v_S$ lifted to PQ breaking scale
 and show that the large hierarchy among the disparate scales is technically natural.
In Section~\ref{sec:axions}, the model  is extended so that the KSVZ-type or the DFSZ-type axion is introduced. The phenomenology of the axion is
outlined. In Section~\ref{sec:conclusion}, we conclude the paper.

%%%%%%%%%%%%%%%%%%%%%%%%%%%%%%%%%%%%%%%%%%%%%%%%%%%%%%%%%%%%%%%%%%%%%%
\section{The model}\label{sec:model}

We briefly recapitulate the model considered in \cite{Baek:2016wml} before discussing the VEVs and their naturalness.
{ Since we promote $U(1)_X$ of \cite{Baek:2016wml,Baek:2018wuo} to $U(1)_{\rm PQ}$ in Section~\ref{sec:axions},
we will call the global symmetry $U(1)_{\rm PQ}$ from now on.
%The scalar fields and a new fermion with their charge assignments under the SM and $U(1)_X$ are shown in Table~\ref{tab:1}.
The charge assignment under the SM and $U(1)_{\rm PQ}$ is shown in Table~\ref{tab:1}. The left-handed SM fermion fields,
which are not listed in Table~\ref{tab:1},  are assigned with PQ-charge 0.}
\begin{widetext}
\begin{center} 
\begin{table}%[tbc]
\renewcommand{\arraystretch}{1.2}
\scriptsize
\begin{tabular}{|c||c|c|c||c|c|c|c||c|c||c|c|}\hline\hline  
&\multicolumn{3}{c||}{Scalar Fields} & \multicolumn{4}{c||}{Fermions} &\multicolumn{2}{c||}{KSVZ} & \multicolumn{2}{c|}{DFSZ} \\\hline
& ~$\Phi_1$~ & ~$\Phi_2$~ & ~$S$~ & ~$u_{Ri}$& ~$d_{Ri}$& ~$e_{Ri}$ & ~$\nu_{Ri}$ &$ \Psi_L$~ & ~$\Psi_R$~ & ~$\Phi_d$ ~ & ~$\Phi_u$~  \\\hline 
$SU(3)_C$ & $\bm{1}$ & $\bm{1}$ & $\bm{1}$ & $\bm{3}$& $\bm{3}$& $\bm{1}$& $\bm{1}$& $\bm{3}$ & $\bm{3}$ & $\bm{1}$ & $\bm{1}$ \\\hline
$SU(2)_L$ & $\bm{2}$  & $\bm{2}$  & $\bm{1}$ & $\bm{1}$& $\bm{1}$& $\bm{1}$& $\bm{1}$  & $\bm{1}$  & $\bm{1}$  & $\bm{2}$ & $\bm{2}$  \\\hline 
$U(1)_Y$ & $\frac12$ & $\frac12$  & $0$& ${2 \over 3}$& $-{1 \over 3}$& $-1$ & $0$ & $e_\Psi$ & $e_\Psi$  & $\frac12$ & $\frac12$ \\\hline
${ U(1)_{\rm PQ}}$ &  $Z_1 (X_1)$ &  $Z_2 (X_2)$   &  $Z_S (X_S)$ 
     &  $Z_{u_R} (X_{u_R})$               &  $Z_{d_R} (X_{d_R})$                   &  $Z_{e_R} (X_{e_R})$                      &  $Z_{\nu_R} (X_{\nu_R})$  
     &  $0$                                                           &  $Z_{\Psi_R} (X_{\Psi_R})$           
     &  $Z_d (X_d)$                                 &  $Z_u (X_u)$    \\\hline
%%%
\end{tabular}
%\caption{Scalar and fermion fields and with their charge assignments under the SM and $U(1)_X$.
\caption{
 Charge assignment under the SM and $U(1)_{\rm PQ}$. { For  the $U(1)_{\rm PQ}$, the $Z$'s ($X$'s) are (effective) PQ charges. 
 The $Z$'s are independent of VEVs, while $X$'s may depend on VEVs.   }
 The explicit values for { $Z$'s and} $X$'s depend on the models and will be shown 
for each model later.
KSVZ (DFSZ) corresponds to the model extension for the KSVZ-(DFSZ-) type axion.
{ We normalize $Z_S=X_S=1$ throughout the paper. We assign $Z_{\Psi_R}=-1$, $Z_d=-1$, and $Z_u=1$.}}
\label{tab:1}
% \end{tiny}
\end{table}
\end{center}
\end{widetext}

\begin{figure}[htbp]
\begin{center}
\includegraphics[width=0.6\textwidth]{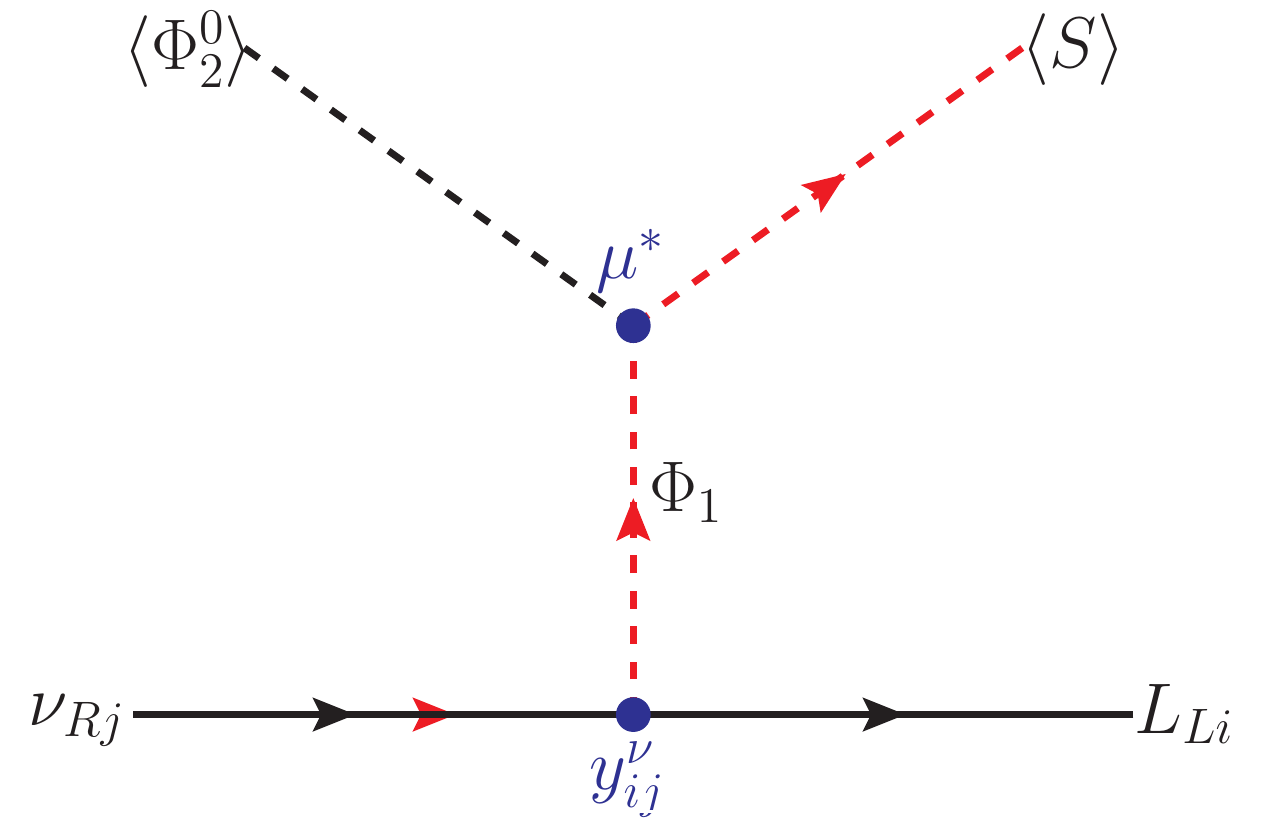}
\caption{Feynman diagram for the neutrino mass generation. The red (black) arrows represent the flow of %$U(1)_X$ 
{ $U(1)_{\rm PQ}$} (lepton number) current.}
\label{fig:neutrino_mass}
\end{center}
\end{figure}

The Yukawa interactions consistent with the SM gauge symmetry and the %$U(1)_X$ 
$U(1)_{\rm PQ}$ symmetry are written as
\eq{
{\cal L} &= -y^u_{ij} \ol{Q}_{Li} \wt{\Phi}_2 u_{Rj}  -y^d_{ij} \ol{Q}_{Li} \Phi_2 d_{Rj}   
              -y^e_{ij} \ol{L}_{Li} \Phi_2 e_{Rj}    -y^\nu_{ij} \ol{L}_{Li} \wt{\Phi}_1 \nu_{Rj}  + h.c.,
\label{eq:model0}
}
where $i,j=1,2,3$ are generation indices, { and  $\wt{\Phi}_{1(2)} = i \sigma^2 \Phi^*_{1(2)}$.}
The scalar potential is given in the form
\eq{
V &= +m_{11}^2 \Phi_1^\dagger \Phi_1 -m_{22}^2 \Phi_2^\dagger \Phi_2 -m_{SS}^2 S^\dagger S
       -(\mu \Phi_1^\dagger \Phi_2 S + h.c) \nl
   &+ \lambda_1 (\Phi_1^\dagger \Phi_1)^2 + \lambda_2 (\Phi_2^\dagger \Phi_2)^2
+ \lambda_3 (\Phi_1^\dagger \Phi_1) (\Phi_2^\dagger \Phi_2) + \lambda_4 (\Phi_1^\dagger \Phi_2) (\Phi_2^\dagger \Phi_1) \nl
&+ \lambda_S (S^\dagger S)^2 +\lambda_{1S} (\Phi_1^\dagger \Phi_1) (S^\dagger S)+\lambda_{2S} (\Phi_2^\dagger \Phi_2) (S^\dagger S).
\label{eq:V}
}
Note that the sign in front of $m_{11}^2$ is ``+". We assume $m_{11}^2>0$, $m_{22}^2>0$, and $m_{SS}^2>0$.
{ We can also take $\mu$ to be positive without loss of generality because any complex phase can be absorbed by
the redefinition of $S$.}
We can decompose the scalars as
\eq{
\Phi_1 = \begin{pmatrix} \phi_1^+ \\ {1 \over \sqrt{2}}(v_1+h_1 +i a_1)\end{pmatrix},\quad
\Phi_2 = \begin{pmatrix} \phi_2^+ \\ {1 \over \sqrt{2}}(v_2+h_2 +i a_2)\end{pmatrix}, \quad
S = {1 \over \sqrt{2}}(v_S + h_S + i a_S).
\label{eq:components}
}
For the vacuum stability we impose the copositivity condition~\cite{Kannike:2012pe}:
\begin{align}
& \lambda_1>0, \quad \lambda_2>0, \quad \lambda_S>0, \nl
& \tilde{\la}_{12} \equiv {1 \over 2} (\la_3+\la_4)+\sqrt{\la_1 \la_2} >0, \nl
& \tilde{\lambda}_{1S} \equiv \frac12 \lambda_{1S} +\sqrt{\lambda_1 \lambda_S} >0, \nl
& \tilde{\lambda}_{2S} \equiv \frac12 \lambda_{2S} +\sqrt{\lambda_2 \lambda_S} >0, \nl
& \sqrt{\la_1 \la_2 \la_3}+\frac12 (\la_3+\la_4) \sqrt{\la_S}+\frac12 \la_{1S} \sqrt{\la_2} +\frac12 \la_{2S} \sqrt{\la_1}+\sqrt{2 \tilde{\la}_{12} \tilde{\la}_{1S} \tilde{\la}_{2S}} >0.
\label{eq:copos}
\end{align}
The conditions in (\ref{eq:copos}) are automatically satisfied when all the $\lambda$'s are positive, which we assume in this paper.
The minimization conditions for the potential along the direction of neutral fields $h_1, h_2$, and $h_S$ are
\begin{align}
\label{eq:v1}
& +2 m_{11}^2 v_1 + 2 \lambda_1 v_1^3 + v_1 (\lambda_{1S} v_S^2 + \lambda_3 v_2^2 + \lambda_4 v_2^2) - \sqrt{2} \mu v_2  v_S =0, \\
& -2 m_{22}^2 v_2 + 2 \lambda_2 v_2^3 + v_2 (\lambda_{2S} v_S^2 + \lambda_3 v_1^2 + \lambda_4 v_1^2) - \sqrt{2} \mu v_1 v_S =0, \\
& -2 m_{SS}^2 v_S + 2 \lambda_S v_S^3 + v_S (\lambda_{1S} v_1^2 + \lambda_{2S} v_2^2 ) - \sqrt{2} \mu v_1 v_2 =0.
\end{align}
From the above coupled equations we solve for VEVs: $v_S$, $v_2$, and $v_1$.
The solution consistent with our assumption, $v_S \gg v_2 \gg v_1$, is given approximately by
\begin{align}
v_S &\approx {m_{SS}  \over \sqrt{ \lambda_S}}, \nl
v_2 &\approx \sqrt{2 m_{22}^2-\lambda_{2S} v_S^2  \over 2 \lambda_2}.
\end{align}
To ensure {\it small} $v_1$, we make an additional assumption,
\begin{align}
\lambda_{1S} v_S^2 + (\lambda_3  + \lambda_4) v_2^2 +2 m_{11}^2 >0,
\end{align}
which can be easily satisfied since $m_{11}^2 \sim {\cal O}(10^{12} \,{\rm GeV})^2$ dominates the left hand side of the inequality.
If $\mu=0$, the only solution for (\ref{eq:v1}) is $v_1=0$. 
For $\mu \neq 0$, the solution is proportional to $\mu$ and is obtained in a good approximation to be
\begin{align}
v_1 &\approx {\sqrt{2} \mu v_2 v_S \over \lambda_{1S} v_S^2  +(\lambda_{3}+\lambda_{4}) v_2^2 + 2 m_{11}^2  } \approx 
{\sqrt{2} \mu v_2  v_S \over 2 m_{11}^2},
\label{eq:tiny_v1}
\end{align} 
where the last relation follows because $ m_{11}^2 \gg |\la_{1S}| v_S^2 \gg |\lambda_{3}+\lambda_{4}| v_2^2 $.
As we mentioned in Section~\ref{sec:introduction}, we require $v_1 \sim  {\cal O}(1 \,{\rm eV})$ to explain small neutrino mass.
We can easily achieve this by taking, for example, $\mu \sim 1 \, {\rm GeV}$,    
when $m_{11} \sim v_S \sim 10^{12} \, {\rm GeV}$. The Feynman diagram for the neutrino mass is shown in Figure~\ref{fig:neutrino_mass}.
The red (black) arrows represent the flow of %$U(1)_X$ 
{ $U(1)_{\rm PQ}$} (lepton number) current. We can see the neutrino mass is effectively given by
\eq{
m_{\nu_{ij}} \sim y^\nu_{ij} \frac{\mu v_2 v_S}{m_{11}^2},
}
from which we can check $v_1$ is the form in (\ref{eq:tiny_v1}).

The superheavy $\Phi_1$ makes the model escape the collider and other low energy bound easily. 
It is noted that $\Phi_1$ is superheavy but its VEV is tiny.
The bottomline is that the main contribution to the masses of $\Phi_1$ components
come from the bare mass term $m_{11}^2$, which makes them superheavy, while $v_1$ generated by the diagram shown in
Figure~\ref{fig:neutrino_mass} can remain tiny. 
The small $\mu$ which makes $v_1$ tiny is technically natural because the symmetry of the Lagrangian (\ref{eq:V}) is enhanced in the $\mu \to 0$ 
limit~\cite{Baek:2016wml}. This can be seen also from the RG equation for $\mu$,
\eq{
{ d \mu(Q) \over d \log Q} &= { 1 \over 8 \pi^2}  \mu \left(\la_{1S} + \la_{2S}+ \la_3+ 2 \la_4\right), 
}
which shows that $\mu=0$ is a fixed point, making small $\mu$ remain small under the change of scale.

We also need to address the large hierarchy between the electroweak scale and the PQ-breaking scale.
The large hierarchy may cause naturalness problem by generating quadratic divergence in the quantum corrections of Higgs mass.
A solution to this problem comes from Poincar\'{e} protection~\cite{Foot:2013hna}. We adopt their paradigm by taking
\begin{align}
\lambda_{2S} \lesssim {\cal O}\left(m_H^2 \over m_S^2\right) \sim 10^{-20} \left(m_H \over 100 \,{\rm GeV}\right)^2 \left(10^{12} \,{\rm GeV} \over m_S\right)^2.
\end{align}
We also take $\lambda_{1S}$ with similar size.
Sending $(\lambda_{1S} ,\lambda_{2S}) \to (0,0)$, the Poincar\'{e} symmetry is enhanced, making the small
$(\lambda_{1S},\lambda_{2S})$ technically natural~\cite{Foot:2013hna}.
We can also check this from the renormalization group running of $\lambda_{1S}$ and $\lambda_{2S}$,
which is given by,
\begin{align}
{d \la_{1S}(Q) \over d \log Q}  &= { 1 \over 8 \pi^2}   \Bigg[2 \la_{1S} (\la_{1S}+3 \la_1+2 \la_S)+  \la_{2S}(2 \la_3+\la_4)\Bigg], \nl
{d \la_{2S}(Q) \over d \log Q} &= { 1 \over 8 \pi^2}   \Bigg[2 \la_{2S} (\la_{2S}+3 \la_2+2 \la_S)+  \la_{1S}(2 \la_3+\la_4)\Bigg],
\end{align}
which reveals $(\lambda_{1S} ,\lambda_{2S}) = (0,0)$ is a fixed point.

\section{Axions}
\label{sec:axions}

Since the VEV $v_S$ can be naturally ${\cal O}(10^{12})$ GeV, we may identify this scale as the breaking scale of the PQ symmetry.
By promoting the $U(1)_X$ symmetry to PQ symmetry $U(1)_{\rm PQ}$, we can introduce axion to solve the strong CP problem.
{ From the charge assignments in Table~\ref{tab:nuTHDM}, the axion field can be defined as~\cite{Srednicki:1985xd}
\eq{
a_{\rm PQ} 
&= {1 \over v^0_{\rm PQ}} \sum_{i=1,2,S}  Z_i v_i a_i\nl
&= {1 \over v^0_{\rm PQ}} \Big( {1 \over 2} v_1 a_1 - {1 \over 2} v_2 a_2 + v_S a_S \Big),
\label{eq:a_PQ}
}
where $v^0_{\rm PQ} =(\sum_i (Z_i v_i)^2)^{1/2}=(v_1^2/4+v_2^2/4+v_S^2)^{1/2}$. However, $a_{\rm PQ}$ is not orthogonal to the Nambu-Goldstone 
boson, $a_Z = (v_1 a_1 + v_2 a_2)/(v_1^2 + v_2^2)^{1/2}$, which is eaten
by the longitudinal component of $Z$ boson. The physical axion field is obtained by orthogonalization:
\eq{
a &= a_{\rm PQ} - \frac{v_1^2 - v_2^2}{2 v^0_{\rm PQ} (v_1^2 + v_2^2)^{1/2}}  a_Z \nl
   &= \frac{1}{v^0_{\rm PQ} (v_1^2 + v_2^2)}(v_1 v_2^2 a_1 - v_1^2 v_2 a_2 + (v_1^2 + v_2^2) v_S a_S).
\label{eq:phys_a}
}
We note that $v^0_{\rm PQ}$ should be changed to $v_{\rm PQ}\equiv(v_S^2+ v_1^2 v_2^2/(v_1^2+v_2^2))^{1/2}$ for $a$ to be
canonically normalized. 
Now we can define the {\em effective} PQ charges, $X_i$ ($i=1,2,S$), so that
\eq{
a = \frac{1}{v_{\rm PQ}} \sum_{i=1,2,S} X_i v_i a_i.
\label{eq:a_X}
}
Then we get
\eq{
X_1 =\frac{v_2^2}{v_1^2+v_2^2}, \quad X_2 =-\frac{v_1^2}{v_1^2+v_2^2}, \quad X_S=1.
\label{eq:E_PQC}
}
}
%The { axion $a$,} Nambu-Goldstone boson of %$U(1)_X$ 
%{ $U(1)_{\rm PQ}$,} appears as
{ Equivalently, we can parametrize the Higgs fields in terms of the physical axion $a$ as}
\eq{
\Phi_1 = {v_1 \over \sqrt{2}} e^{i a X_1 /v_{\rm PQ}}
\begin{pmatrix}0 \\ 1 \end{pmatrix}, \quad
\Phi_2 = {v_2 \over \sqrt{2}} e^{i a X_2/v_{\rm PQ}}
\begin{pmatrix}0 \\ 1 \end{pmatrix}, \quad
S ={v_S \over \sqrt{2}} e^{i a X_S /v_{\rm PQ}},
\label{eq:nonlinear_axion}
}
where $X_1, X_2$, and $X_S$ are their { effective} PQ charges. 
{ We will calculate the $X$'s again and show that they agree with (\ref{eq:E_PQC}).}
We take the normalization $X_S \equiv 1$.
The $\mu$ term in (\ref{eq:V}) dictates
\eq{
-X_1 + X_2 + X_S =0.
}
Since $a$ should not mix with the $Z$-boson, we get a condition
\eq{
X_1 v_1^2 + X_2 v_2^2 =0.
\label{eq:Z-axion}
}
{ We refer the reader to Appendix~\ref{sec:appendix} for details of the above relation.
}
Then we get
%\eq{
%X_1 = \cos^2\be_\nu, \quad X_2 = -\sin^2\be_\nu,
%}
%where $\be_\nu$ is defined so that $\tan\be_\nu =v_1/v_2$.
\eq{
 X_1 = {v_2^2 \over v_1^2+v_2^2}, \quad X_2 = -{v_1^2 \over v_1^2+v_2^2}.
\label{eq:E_PQC2}
}
The SM electroweak vacuum is obtained to be $v_{\rm ew}\equiv\sqrt{v_1^2+v_2^2} \simeq 246$ GeV.
{ The result with the normalization $X_S =1$ agrees with (\ref{eq:E_PQC}).}
Expanding (\ref{eq:nonlinear_axion}) up to linear order, we can identify the axion field $a$ in terms of $(a_1, a_2, a_S)$,
\eq{
a &={1 \over v_{\rm PQ}}(X_1 v_1 a_1+X_2 v_2 a_2+X_S v_S a_S) \nl
&={1 \over v_{\rm PQ} (v_1^2+v_2^2)} (v_1 v_2^2 a_1 -v_1^2 v_2 a_2 + v_{\rm ew}^2 v_S a_S){,}
}
{ which agrees with (\ref{eq:phys_a}) up to the overall constant.}
From normalization condition we can express  $v_{\rm PQ}$ as
\eq{
v_{\rm PQ}=\left((X_1 v_1)^2 +(X_2 v_2)^2 +(X_S v_S)^2\right)^{1/2}
={1 \over v_1^2+v_2^2}\left((v_1 v_2^2)^2 +(v_1^2 v_2)^2 +(v_{\rm ew}^2 v_S)^2\right)^{1/2}{,}
}
{ which is exactly what we obtained below (\ref{eq:a_X}).}
From (\ref{eq:model0}) the { effective} charges for the other fields read
\eq{
X_{u_R} = X_2 ={ -{v_1^2 \over  v_1^2+v_2^2}},
 \quad X_{d_R} = X_{e_R} =- X_2 ={ {v_1^2 \over  v_1^2+v_2^2}}, 
 \quad X_{\nu_R} = X_1 = { {v_2^2  \over v_1^2+v_2^2}}. 
 \label{eq:X_nuTHDM}
}

\begin{widetext}
\begin{center} 
\begin{table}[t]
%\begin{tiny}
{
{\renewcommand{\arraystretch}{1.2}
\begin{tabular}{|c||c|c|c||c|c|c|c|}\hline\hline  
         &\multicolumn{3}{c||}{Scalar Fields} 
         & \multicolumn{4}{c|}{Fermions} \\\cline{2-8}
                & ~$\Phi_1 (X_1)$~            & ~$\Phi_2(X_2)$~                      & ~$S(X_S)$~ 
                & ~$u_{Ri} (X_{u_R})$        & ~$d_{Ri} (X_{d_R})$                & ~$e_{Ri}(X_{e_R})$                       & ~$\nu_{Ri}(X_{\nu_R})$  \\\hline 
~$U(1)_{\rm PQ}$~ 
          &  { ${1 \over 2}\left({v_2^2 \over v_1^2+v_2^2}\right)$}           & { $-{1 \over 2}\left(-{v_1^2 \over v_1^2+v_2^2}\right)$}          & { 1 (1)}
          &  { $-{1 \over 2}\left(-{v_1^2 \over v_1^2+v_2^2}\right)$}           &  { ${1 \over 2}\left({v_1^2 \over v_1^2+v_2^2}\right)$}           &  { ${1 \over 2}\left({v_1^2 \over v_1^2+v_2^2}\right)$}           &  { ${1 \over 2}\left({v_2^2 \over v_1^2+v_2^2}\right)$} 
            \\\hline
%%%
\end{tabular}
}
}
%\caption{Scalar and fermion fields and with their charge assignments under the SM and $U(1)_X$.
\caption{
 The non-vanishing  $U(1)_{\rm PQ}$ { (effective)} charges for the $\nu$THDM described by (\ref{eq:model0}) and (\ref{eq:V}).}
\label{tab:nuTHDM}
% \end{tiny}
\end{table}
\end{center}
\end{widetext}
%%%%
{The non-vanishing PQ-charges for the $\nu$THDM described by (\ref{eq:model0}) and (\ref{eq:V}) are summarized in Table~\ref{tab:nuTHDM}.}
Since the PQ charges of $u_{Ri}$ and $d_{Ri}$ add to zero, the QCD anomaly cancels, and the Nambu-Goldstone boson
$a$ cannot be QCD axion candidate in the model (\ref{eq:model0}). 

We can extend the model to make $a$ an axion to solve the strong CP problem in two ways.
One method is to introduce a pair of heavy vector-like quarks $\Psi_{L,R}$ which couples to $S$ and weighs PQ-breaking scale, where
$a$ becomes KSVZ-type axion~\cite{Kim:1979if,Shifman:1979if}.
The other one is by replacing $\Phi_2$ with a pair of Higgs doublet $\Phi_{u,d}$, where $\Phi_{u(d)}$ couples to $S$ and $u(d)$-type quarks,
making $a$ DFSZ-type axion~\cite{Dine:1981rt,Zhitnitsky:1980tq}.
The charge assignments for the two possible scenarios are shown in Table~\ref{tab:1} { with tabs KSVZ and DFSZ}.
In both scenarios the seesaw-like relation (\ref{eq:tiny_v1}) and axion decay constant in (\ref{eq:KSVZ_ma}) and (\ref{eq:fa_DFSZ})
share the PQ-breaking parameter $v_S$ in common, showing the interplay between neutrino mass and axion.
As a consequence, neutrino experiments can constrain the parameter space of the axion model, and vice versa~\cite{Peinado:2019mrn}.

\subsection{KSVZ-type axion}

{ In the KSVZ-type scenario for axion we introduce heavy vector-like quarks in addition to fields shown in Table~\ref{tab:nuTHDM}.}
%In the KSVZ-type scenario 
We have Yukawa coupling for $\Psi$: 
\eq{
\Delta {\cal L} = -y_\Psi \overline{\Psi}_L S \Psi_R + h.c.
\label{eq:KSVZ}
}
Then the  { (effective)} PQ-charge of $\Psi_R$ is fixed: { $Z_{\Psi_R}=-Z_S = -1$ ($X_{\Psi_R}=-X_S = -1$)}, assuming { $Z_{\Psi_L}=0$ ($X_{\Psi_L}=0$)}.
The Yukawa interactions and the scalar potential  shown in (\ref{eq:model0}) and (\ref{eq:V})  remain the same.
As a consequence, the equations from (\ref{eq:nonlinear_axion}) to (\ref{eq:X_nuTHDM}) hold also in this model without any change. 
The PQ charges for this model are collected in Table~\ref{tab:KSVZ}
%%%%%
\begin{widetext}
\begin{center} 
\begin{table}[tb]
\renewcommand{\arraystretch}{1.2}
\scriptsize
\begin{tabular}{|c||c|c|c||c|c|c|c||c|c|}\hline\hline  
&\multicolumn{3}{c||}{Scalar Fields} & \multicolumn{4}{c||}{Fermions} &\multicolumn{2}{c|}{KSVZ}\\\cline{2-10}
      & ~$\Phi_1 (X_1)$~ & ~$\Phi_2(X_2)$~ & ~$S(X_S)$~ 
      & ~$u_{Ri} (X_{u_R})$& ~$d_{Ri} (X_{d_R})$& ~$e_{Ri}(X_{e_R})$ & ~$\nu_{Ri}(X_{\nu_R})$  
      & ~$\Psi_L(X_{\Psi_L})$~ & ~$\Psi_R (X_{\Psi_R})$~\\\hline
~$U(1)_{\rm PQ}$~ 
                &  { ${1 \over 2}\left({v_2^2 \over v_1^2+v_2^2}\right)$}           & { $-{1 \over 2}\left(-{v_1^2 \over v_1^2+v_2^2}\right)$}          & { 1 (1)}
          &  { $-{1 \over 2}\left(-{v_1^2 \over v_1^2+v_2^2}\right)$}           &  { ${1 \over 2}\left({v_1^2 \over v_1^2+v_2^2}\right)$}           &  { ${1 \over 2}\left({v_1^2 \over v_1^2+v_2^2}\right)$}           &  { ${1 \over 2}\left({v_2^2 \over v_1^2+v_2^2}\right)$} 
           & { $0 (0)$}    & { $-1 (-1)$ }  \\\hline
%%%
\end{tabular}
%\caption{Scalar and fermion fields and with their charge assignments under the SM and $U(1)_X$.
\caption{
The $U(1)_{\rm PQ}$ { (effective)} charges for the KSVZ-type model described by (\ref{eq:model0}),  (\ref{eq:V}) and (\ref{eq:KSVZ}). The PQ charges
for the fields not listed here all vanish.}
\label{tab:KSVZ}
\end{table}
\end{center}
\end{widetext} 
Independently of %the angle $\be_\nu$ 
{ the ratio $v_1/v_2$} the axion mass is given by~\cite{diCortona:2015ldu}
\eq{
m_a &={ f_\pi m_\pi \over {v_{\rm PQ}}} {\sqrt{m_u m_d} \over m_u+m_d}\simeq 5.7 \, {\rm \mu eV} \left(10^{12} \,{\rm GeV} \over f_a\right),
\label{eq:KSVZ_ma}
}
where in the last equality we identified the axion decay constant $f_a \equiv v_{\rm PQ}$. 
The axion-photon coupling is
\eq{
{\cal L}_{a\gamma\gamma}=-\Big[6 e_\Psi^2-\frac{2(4+z)}{3(1+z)}\Big] {a \over v_{\rm PQ}} \frac{e^2}{32\pi^2} F^{\mu\nu} \widetilde{F}_{\mu\nu}
%=\Big[6 e_\Psi^2-1.92\Big] {a \over \cancelto{f_a}{v_{\rm PQ}}} \frac{\alpha_{\rm em}}{2\pi \cancelto{}{f_a}} \boldsymbol{E}\cdot\boldsymbol{B}.
=\Big[6 e_\Psi^2-1.92\Big] \frac{\alpha_{\rm em}}{2\pi f_a} \,a\, \boldsymbol{E}\cdot\boldsymbol{B} 
\label{eq:KSVZ_arr}
}
The $a\gamma\gamma$-coupling is also independent of %$\beta_\nu$
{ the ration $v_1/v_2$}. The results (\ref{eq:KSVZ_ma}) and (\ref{eq:KSVZ_arr}) are in exact agreement with those of the original KSVZ 
model~\cite{Kim:1979if,Shifman:1979if}. The difference comes in the couplings of the axion to electron and neutrinos.
In general the axion coupling to fermions is written as~\cite{Tanabashi:2018oca},
\eq{
{\cal L}_{aff} = -\frac{C_f}{2 f_a} \partial_\mu a \overline{f} \gamma^\mu \gamma_5 f.
}
The tree level axion coupling to electrons (neutrinos) is %$C_e = \sin^2\be_\nu \; (C_\nu = \cos^2\be_\nu)$, 
{ $C_e = v_1^2/(v_1^2+v_2^2)  \; (C_\nu = v_2^2/(v_1^2+v_2^2))$}, 
while they both vanish in the KSVZ model.
However, the fact that %$\beta_\nu \lll 1$ 
{ $v_1 \ll v_2$} makes the $C_e$ coupling practically indistinguishable from that in the KSVZ model. The axion coupling to
neutrinos $C_\nu \approx 1$ and in principle can be probed through experiments such as neutrino oscillation~\cite{Huang:2018cwo}. 

\subsection{DFSZ-type axion}
We can also extend the Higgs sector to introduce a QCD axion~\cite{Dine:1981rt,Zhitnitsky:1980tq}. We replace $\Phi_2$ with $\Phi_u$ and $\Phi_d$, where
$\Phi_u$ couples only to the up-type quarks and $\Phi_d$ couples only to the down-type quarks.
%Their charge assignments are shown in Table~\ref{tab:1}.
The charged-leptons can couple either to $\Phi_d$ (type-II) or to $\Phi_u$ (flipped).
%We assign the PQ-charge $X_u$ ($X_d$) to $\Phi_u$ ($\Phi_d$). 
%The PQ-charge of $\Phi_1$ is still $X_1$.
As in the original DFSZ model, we introduce
\eq{
  \Phi_u^\dagger \Phi_d S^2 + h.c.
\label{eq:phi_udS2}
}
term to the Lagrangian.
Now the $\mu$-term in (\ref{eq:V}) can take the form of either
\eq{
-\mu \Phi_1^\dagger \Phi_d S + h.c.,
\label{eq:phi_d}
}
or
\eq{
-\mu \Phi_1^\dagger \Phi_u S + h.c.,
\label{eq:phi_u}
}
but not both\footnote{
If both terms are introduced into the Lagrangian simultaneously, 
the equality of $q_u$, $q_d$ makes the QCD anomaly cancel and $a$ cannot play the role of axion.}.
The introduction of (\ref{eq:phi_udS2}) and (\ref{eq:phi_d}) (or  (\ref{eq:phi_u})) does not spoil the
smallness of $v_1$ in (\ref{eq:tiny_v1}) and the naturalness arguments presented in Section~\ref{sec:model}.
% Since the case (\ref{eq:phi_u})
%can be obtained simply by exchanging $d \leftrightarrow u$ from the case of (\ref{eq:phi_d}),
%we concentrate only on (\ref{eq:phi_d}). 
{ The PQ charges are assigned as in Table~\ref{tab:DFSZ}. And the $X$'s are calculated as follows\footnote{We also checked that
the method using (\ref{eq:a_PQ}) and (\ref{eq:phys_a}) gives the same results.}.}
The (\ref{eq:phi_udS2}) gives a relation,
\eq{
-X_u + X_d +2 X_S =0,
\label{eq:udS2}
}
where we will take the normalization $X_S=1$ as in the KSVZ scenario. 
From (\ref{eq:phi_d}) ((\ref{eq:phi_u})) the { effective} PQ-charges should satisfy
\eq{
-X_1 + X_d +X_S =0, \quad (-X_1 + X_u +X_S =0).
\label{eq:1dS}
}
The axion does not mix with the SM $Z$-boson, if
\eq{
X_1 v_1^2 + X_d v_d^2 + X_u v_u^2 =0,
\label{eq:no_mixing}
}
where $\langle \Phi^0_{d(u)}\rangle=v_{d(u)}/\sqrt{2}$. Since $\sqrt{v_u^2+v_d^2+v_1^2}\equiv v_{\rm ew} \simeq 246$ GeV,
$v_{d(u)}$ is at most at the electroweak scale.
If the kinetic term of the axion is canonically normalized, we get
\eq{
X_1^2 v_1^2 + X_d^2 v_d^2 + X_u^2 v_u^2 +X_S^2 v_S^2 =v_{\rm PQ}^2.
\label{eq:vPQ_DFSZ}
}

Now we can solve $X_1$, $X_d$ and $X_u$ from (\ref{eq:udS2}), (\ref{eq:1dS}), and (\ref{eq:no_mixing}), setting $X_S=1$:
for (\ref{eq:phi_d}) we get
\eq{
X_1 &= \frac{v_d^2-v_u^2}{v_1^2+v_d^2+v_u^2},\quad X_d =  -\frac{v_1^2+2v_u^2}{v_1^2+v_d^2+v_u^2},
\quad X_u =  \frac{v_1^2+2v_d^2}{v_1^2+v_d^2+v_u^2},
\label{eq:sol1}
}
and for (\ref{eq:phi_u}) we get
\eq{
X_1 &= \frac{3 v_d^2+v_u^2}{v_1^2+v_d^2+v_u^2},\quad X_d =  -\frac{3 v_1^2+2v_u^2}{v_1^2+v_d^2+v_u^2},
\quad X_u =  -\frac{v_1^2-2v_d^2}{v_1^2+v_d^2+v_u^2}.
\label{eq:sol2}
}
Then the PQ-breaking scale $v_{\rm PQ}$ can be determined by (\ref{eq:vPQ_DFSZ}).
Inserting (\ref{eq:sol1}) and (\ref{eq:sol2}), we get
\eq{
v_{\rm PQ}^2 = v_S^2+\frac{ v_1^2 (v_d^2+v_u^2)+4 v_d^2 v_u^2}{v_1^2+v_d^2+v_u^2},\;  \text{and} \quad
v_{\rm PQ}^2 = v_S^2+\frac{v_1^2 (9 v_d^2+v_u^2)+4 v_d^2 v_u^2}{v_1^2+v_d^2+v_u^2}, 
}
respectively.
{ In this model the Yukawa interactions in (\ref{eq:model0}) are written as
\eq{
{\cal L} &= -y^u_{ij} \ol{Q}_{Li} \wt{\Phi}_u u_{Rj}  -y^d_{ij} \ol{Q}_{Li} \Phi_d d_{Rj}   
              -y^e_{ij} \ol{L}_{Li} \Phi_q e_{Rj}    -y^\nu_{ij} \ol{L}_{Li} \wt{\Phi}_1 \nu_{Rj}  + h.c.,
\label{eq:Yuk_DFSZ}
}
where $q=d (u)$ for type-II (flipped) scenario for the lepton couplings to Higgs. From (\ref{eq:Yuk_DFSZ}) we get
\eq{
X_{u_R} = X_u, \quad X_{d_R} = -X_d, \quad X_{\nu R}=X_1.
}
Depending on the scenario, the $X_{e_R}$ is 
\eq{
X_{e_R} = -X_q = \left\{ \begin{array}{l} -X_d \;\; \text{(type-II)} \\ -X_u \;\; \text{(flipped).}\end{array} \right. 
}
So we have four possible models whose PQ-charges are summarized in Table~\ref{tab:DFSZ}.
}

%%%%%
\begin{widetext}
\begin{center} 
\begin{table}[tb]
{
{\renewcommand{\arraystretch}{1.2}
\tiny
{
\begin{tabular}{|c|c||c|c||c|c|c|c||c|c|}\hline\hline  
\multicolumn{2}{|c||}{} &\multicolumn{2}{c||}{Scalar Fields} & \multicolumn{4}{c||}{Fermions} &\multicolumn{2}{c|}{DFSZ}\\\cline{3-10}
\multicolumn{2}{|c||}{} & ~$\Phi_1 (X_1)$~ &  ~$S(X_S)$~ & ~$u_{Ri} (X_{u_R})$& ~$d_{Ri} (X_{d_R})$& ~$e_{Ri}(X_{e_R})$ & ~$\nu_{Ri}(X_{\nu_R})$  & ~$\Phi_d (X_d)$~ & ~$\Phi_u (X_u)$~\\\hline
\multirow{2}{*}{(\ref{eq:phi_d})} 
     &type-II           
     & ~$ 0 \left(\frac{v_d^2-v_u^2}{v_1^2+v_d^2+v_u^2}\right)$~                 &  $1(1)$ 
     &  $1 \lt(\frac{v_1^2+2v_d^2}{v_1^2+v_d^2+v_u^2} \rt)$                 & $1\lt(\frac{v_1^2+2v_u^2}{v_1^2+v_d^2+v_u^2}\rt)$                        & $1\lt(\frac{v_1^2+2v_u^2}{v_1^2+v_d^2+v_u^2}\rt)$               & ~$0\lt(\frac{v_d^2-v_u^2}{v_1^2+v_d^2+v_u^2}\rt)$~ 
     &  $-1 \left(-\frac{v_1^2+2v_u^2}{v_1^2+v_d^2+v_u^2} \right)$               & $1 \left(\frac{v_1^2+2v_d^2}{v_1^2+v_d^2+v_u^2}\right)$ \\
 \cline{2-10}
     & flipped& ~$0\lt(\frac{v_d^2-v_u^2}{v_1^2+v_d^2+v_u^2}\rt)$     &  $1\lt(1\rt)$~ 
     & $1 \lt(\frac{v_1^2+2v_d^2}{v_1^2+v_d^2+v_u^2}\rt)$             & $1\lt(\frac{v_1^2+2v_u^2}{v_1^2+v_d^2+v_u^2}\rt)$                     & $-1\lt(-\frac{v_1^2+2v_d^2}{v_1^2+v_d^2+v_u^2}\rt)$                       & ~$0\lt(\frac{v_d^2-v_u^2}{v_1^2+v_d^2+v_u^2}\rt)$~ 
     &  $-1\lt(-\frac{v_1^2+2v_u^2}{v_1^2+v_d^2+v_u^2} \rt)$              & $1\lt(\frac{v_1^2+2v_d^2}{v_1^2+v_d^2+v_u^2}\rt)$ \\\hline
  \multirow{2}{*}{(\ref{eq:phi_u})} 
     & type-II 
     & ~$2 \lt(\frac{3v_d^2+v_u^2}{v_1^2+v_d^2+v_u^2} \rt)$~              &  $1(1)$~ 
     & $1\lt(-\frac{v_1^2-2v_d^2}{v_1^2+v_d^2+v_u^2}\rt)$                  & $1\lt(\frac{3v_1^2+2v_u^2}{v_1^2+v_d^2+v_u^2}\rt)$                  & $1\lt(\frac{3v_1^2+2v_u^2}{v_1^2+v_d^2+v_u^2}\rt)$                 & ~$2\lt(\frac{3v_d^2+v_u^2}{v_1^2+v_d^2+v_u^2}\rt)$~ 
     &  $-1\lt(-\frac{3v_1^2+2v_u^2}{v_1^2+v_d^2+v_u^2}\rt)$             & $1\lt(-\frac{v_1^2-2v_d^2}{v_1^2+v_d^2+v_u^2}\rt)$ \\
 \cline{2-10}
     &flipped  
     & ~$2\lt(\frac{3v_d^2+v_u^2}{v_1^2+v_d^2+v_u^2}\rt)$                & $1(1)$~ 
     & $1\lt(-\frac{v_1^2-2v_d^2}{v_1^2+v_d^2+v_u^2}\rt)$                 & $1\lt(\frac{3v_1^2+2v_u^2}{v_1^2+v_d^2+v_u^2}\rt)$                  & $-1\lt(\frac{v_1^2-2v_d^2}{v_1^2+v_d^2+v_u^2}\rt)$                   & ~$2\lt(\frac{3v_d^2+v_u^2}{v_1^2+v_d^2+v_u^2}\rt)$~ 
     &  $-1\lt(-\frac{3v_1^2+2v_u^2}{v_1^2+v_d^2+v_u^2}\rt)$             & $1\lt(-\frac{v_1^2-2v_d^2}{v_1^2+v_d^2+v_u^2}\rt)$ \\
     \hline
%%%
\end{tabular}}
}
}
%\caption{Scalar and fermion fields and with their charge assignments under the SM and $U(1)_X$.
\caption{
The $U(1)_{\rm PQ}$ {(effective)} charges for the DFSZ-type model described by (\ref{eq:phi_udS2}),  (\ref{eq:phi_d}) (or (\ref{eq:phi_u})), and (\ref{eq:Yuk_DFSZ}). The PQ charges
for the fields not listed here all vanish.}
\label{tab:DFSZ}
\end{table}
\end{center}
\end{widetext}
%%%%%

We can identify the axion decay constant $f_a$ as
\eq{
f_a \equiv \frac{v_{\rm PQ}}{2 N_g},
\label{eq:fa_DFSZ}
}
where $N_g=3$ is the number of generations.
The axion mass is obtained to be
\eq{
m_a=  \frac{f_\pi m_\pi}{f_a} \frac{\sqrt{m_u m_d}}{m_u+m_d} \simeq 5.7 \, {\rm \mu eV} \left(10^{12} \,{\rm GeV} \over f_a\right).
}
The axion-photon coupling reads
\eq{
{\cal L}_{a\gamma\gamma}
=\Big[{E \over N }-\frac{2}{3}\frac{4+z}{1+z}\Big]  \frac{\alpha_{\rm em}}{2\pi f_a} \,a\, \boldsymbol{E}\cdot\boldsymbol{B},
}
where $E/N=8/3 (2/3)$ for type-II (flipped) model.

The axion coupling constants to electrons and neutrinos are obtained to be
\eq{
C_e &= \left\{ \begin{array}{c}  \frac{2v_u^2+v_1^2}{6(v_d^2+v_u^2+v_1^2)} \simeq {1 \over 3} \sin^2\be \quad\text{(type-II)} 
\\  -\frac{2v_d^2+v_1^2}{6(v_d^2+v_u^2+v_1^2)} \simeq -{1 \over 3} \cos^2\be \quad \text{(flipped)},
\end{array}\right. \nl
C_\nu &= -\frac{v_d^2-v_u^2}{6(v_d^2+v_u^2+v_1^2)} \simeq -{1 \over 6} \cos 2\be,
}
for the case of (\ref{eq:phi_d}) and
\eq{
C_e &= \left\{ \begin{array}{c}  \frac{2v_u^2+3v_1^2}{6(v_d^2+v_u^2+v_1^2)} \simeq {1 \over 3} \sin^2\be \quad\text{(type-II)} 
\\  -\frac{2v_d^2-v_1^2}{6(v_d^2+v_u^2+v_1^2)} \simeq -{1 \over 3} \cos^2\be \quad \text{(flipped)},
\end{array}\right. \nl
C_\nu &= -\frac{3v_d^2+v_u^2}{6(v_d^2+v_u^2+v_1^2)} \simeq -{1 \over 6} (2+\cos 2\be),
}
for the case of (\ref{eq:phi_u}).
In the last approximate equality we neglected small $v_1$ and defined $\tan\be = v_u/v_d$.
The value $C_e$ for type-II model agrees approximately with that of DFSZ model.

\section{Conclusions}
\label{sec:conclusion}

We proposed a model in which neutrino mass generation and axion are connected. This shows that there may be a 
strong interplay among neutrino mass, strong CP puzzle, and dark matter.
In the model the Peccei-Quinn symmetry is broken by the VEV of a singlet scalar field $S$.
The PQ symmetry also forbids the mass term for the right-handed neutrinos. As a consequence the resulting neutrinos
are Dirac type. The VEV $v_1$, which gives the Dirac neutrino mass by the relation $m_\nu = y_\nu v_1/\sqrt{2}$,
is $\sim {\cal O}(1 \, {\rm eV})$, while the Yukawa coupling $y_\nu$ can be of order one.
The neutrino mass generation is depicted by the diagram in Figure~\ref{fig:neutrino_mass}.
We have shown that the small $v_1$ and the large hierarchy $v_1 \ll v_{\rm ew} \ll v_{\rm PQ}$ are technically natural
by symmetry arguments.

The axion can be realized as either KSVZ-type or DFSZ-type. In both cases the smallness of $v_1$ compared with $v_{\rm ew}{ (\simeq 246\;{\rm GeV})}$
makes the phenomenology of the model almost indistinguishable from the original KSVZ or DFSZ models.
The axion coupling to neutrinos are new to our model, and may be probed in the future neutrino oscillation or axion search experiments.

{\appendix
\section{The identification of axion}
\label{sec:appendix}
To identify the axion among $(a_1, a_2, a_S)$, let us consider the consequences of the symmetries of the potential $V$: 
$SU(2)_L \times U(1)_Y \times U(1)_{\rm PQ}$. The potential $V$ should be invariant under infinitesimal transformations
\eq{
\Phi_1 &\to \Phi'_1 =\left(1+ i \alpha^a {\sigma^a \over 2}+ i \beta Y + i \gamma X_1 \right) \Phi_1, \nl 
\Phi_2 &\to \Phi'_2 =\left(1+ i \alpha^a {\sigma^a \over 2}+ i \beta Y + i \gamma X_2 \right) \Phi_2, \nl 
S &\to S'=\left(1+ i \gamma X_S \right) S,
\label{eq:V_sym}
}
where $Y=1/2$ and $\alpha^a, \beta, \gamma$ are infinitesimal transformation parameters for $SU(2)_L$, $U(1)_Y$, and $U(1)_{\rm PQ}$,
respectively.
The invariance of $V$ under the transformation (\ref{eq:V_sym}), {\it i.e.}
\eq{
V(\Phi_1,\Phi_2,S)=V(\Phi'_1,\Phi'_2,S'),
}
leads to
\eq{
{\cal M}^2 
\begin{pmatrix} 
\left(\frac{-\alpha^3+\beta}{2}+\gamma X_1\right) v_1 \\
\left(\frac{-\alpha^3+\beta}{2}+\gamma X_2\right) v_2 \\
\gamma X_S v_S
\end{pmatrix}=0,
\label{eq:Goldstone_thm}
}
where ${\cal M}^2$ is the pseudo-scalar mass matrix in the basis $(a_1, a_2, a_S)$: ${\cal M}^2_{kl}=\partial^2 V/ \partial a_k \partial a_l |_0$ $(k,l=1,2,S)$.
The equation (\ref{eq:Goldstone_thm}) is the manifestation of the Goldstone's theorem in our model. It reveals two Goldstone bosons: the
one proportional to $(v_1, v_2, 0)$ is the longitudinal component of $Z$ boson and the one proportional to $(X_1 v_1, X_2 v_2, X_S v_S)$
is the axion. Since the two Goldstones should be orthogonal to each other, we require
\eq{
X_1 v_1^2 + X_2 v_2^2 =0,
}
which agrees with (\ref{eq:Z-axion}). The remaining third direction proportional to $(-v_2 v_S, v_1 v_S, v_1 v_2)$ is the massive pseudo-scalar boson.

The relation (\ref{eq:Z-axion}) can be seen also from the kinetic energy terms,
\eq{
{\cal L}_{\rm kin} = |{\cal D}_\mu \Phi_1|^2 +|{\cal D}_\mu \Phi_2|^2 +|{\cal D}_\mu S|^2,
\label{eq:kin}
}
where ${\cal D}_\mu = \partial_\mu + i g/\sqrt{2} (W^+_\mu T^+ +W^-_\mu T^-)  + i g_Z Z_\mu (T^3 - Q sw^2) + i e A_\mu Q$ is
the covariant derivative. Inserting (\ref{eq:components}) into (\ref{eq:kin}), we get a mixing term between $Z$ boson and Goldstone boson:
\eq{
{\cal L}_{\rm mix} = - {g_Z \over 2} Z^\mu \partial_\mu ( v_1 a_1 + v_2 a_2),
}
which shows the Goldstone boson eaten by the $Z$ boson is the one proportional to $(v_1, v_2, 0)$.
Now by inserting (\ref{eq:nonlinear_axion}) into (\ref{eq:kin}), we get a mixing term between $Z$ boson and axion
\eq{
{\cal L}_{\rm mix} =- {g_Z \over 2} Z^\mu \partial_\mu a ( X_1 v_1^2 + X_2 v_2^2),
}
which should vanish. This again confirms  (\ref{eq:Z-axion}).

Alternatively, we can consider the PQ current,
\eq{
J_{\rm PQ}^\mu &= -i ( X_1 \Phi_1^\dagger \partial^\mu \Phi_1 +  X_2 \Phi_2^\dagger \partial^\mu \Phi_2 + X_S S^\dagger \partial^\mu S) +\cdots \nl
&= {1 \over 2} \partial^\mu (X_1 v_1 a_1 + X_2 v_2 a_2+ X_S v_S a_S) + \cdots,  
}
where the ellipses denote terms which do not contain the scalar fields. Since the PQ current should not create or destroy the 
Goldstone boson eaten by the $Z$ boson, we require~\cite{Dine:1981rt}
\eq{
\langle 0 | J_{\rm PQ}^\mu |a_Z \rangle \equiv {1 \over \sqrt{v_1^2+v_2^2}} \langle 0 | J_{\rm PQ}^\mu \left( v_1 |a_1 \rangle + v_2  |a_2 \rangle \right) =0.
}
The above equation again yields,  
\eq{
X_1 v_1^2 + X_2 v_2^2 =0,
}
which agrees with (\ref{eq:Z-axion}).
}

\begin{acknowledgements}
This work was supported by the National Research Foundation of Korea(NRF) grant funded by the Korea government(MSIT) 
(Grant No. NRF-2018R1A2A3075605).
\end{acknowledgements}

%%%%%%%%%%%%%%%%%%%%%%%%%%%%%%%%%%%%%%%%%%%%%%%%%%%%%%%%%%%%%%%%%%%%%%
\bibliographystyle{apsrev4-1}
\bibliography{NPQ_4}

\begin{thebibliography}{31}%
\makeatletter
\providecommand \@ifxundefined [1]{%
 \@ifx{#1\undefined}
}%
\providecommand \@ifnum [1]{%
 \ifnum #1\expandafter \@firstoftwo
 \else \expandafter \@secondoftwo
 \fi
}%
\providecommand \@ifx [1]{%
 \ifx #1\expandafter \@firstoftwo
 \else \expandafter \@secondoftwo
 \fi
}%
\providecommand \natexlab [1]{#1}%
\providecommand \enquote  [1]{``#1''}%
\providecommand \bibnamefont  [1]{#1}%
\providecommand \bibfnamefont [1]{#1}%
\providecommand \citenamefont [1]{#1}%
\providecommand \href@noop [0]{\@secondoftwo}%
\providecommand \href [0]{\begingroup \@sanitize@url \@href}%
\providecommand \@href[1]{\@@startlink{#1}\@@href}%
\providecommand \@@href[1]{\endgroup#1\@@endlink}%
\providecommand \@sanitize@url [0]{\catcode `\\12\catcode `\$12\catcode
  `\&12\catcode `\#12\catcode `\^12\catcode `\_12\catcode `\%12\relax}%
\providecommand \@@startlink[1]{}%
\providecommand \@@endlink[0]{}%
\providecommand \url  [0]{\begingroup\@sanitize@url \@url }%
\providecommand \@url [1]{\endgroup\@href {#1}{\urlprefix }}%
\providecommand \urlprefix  [0]{URL }%
\providecommand \Eprint [0]{\href }%
\providecommand \doibase [0]{http://dx.doi.org/}%
\providecommand \selectlanguage [0]{\@gobble}%
\providecommand \bibinfo  [0]{\@secondoftwo}%
\providecommand \bibfield  [0]{\@secondoftwo}%
\providecommand \translation [1]{[#1]}%
\providecommand \BibitemOpen [0]{}%
\providecommand \bibitemStop [0]{}%
\providecommand \bibitemNoStop [0]{.\EOS\space}%
\providecommand \EOS [0]{\spacefactor3000\relax}%
\providecommand \BibitemShut  [1]{\csname bibitem#1\endcsname}%
\let\auto@bib@innerbib\@empty
%</preamble>
\bibitem [{\citenamefont {Ma}(2001)}]{Ma:2000cc}%
  \BibitemOpen
  \bibfield  {author} {\bibinfo {author} {\bibfnamefont {E.}~\bibnamefont
  {Ma}},\ }\href {\doibase 10.1103/PhysRevLett.86.2502} {\bibfield  {journal}
  {\bibinfo  {journal} {Phys. Rev. Lett.}\ }\textbf {\bibinfo {volume} {86}},\
  \bibinfo {pages} {2502} (\bibinfo {year} {2001})},\ \Eprint
  {http://arxiv.org/abs/hep-ph/0011121} {arXiv:hep-ph/0011121 [hep-ph]}
  \BibitemShut {NoStop}%
%%CITATION = HEP-PH/0011121;%%
\bibitem [{\citenamefont {Grimus}\ \emph {et~al.}(2009)\citenamefont {Grimus},
  \citenamefont {Lavoura},\ and\ \citenamefont {Radovcic}}]{Grimus:2009mm}%
  \BibitemOpen
  \bibfield  {author} {\bibinfo {author} {\bibfnamefont {W.}~\bibnamefont
  {Grimus}}, \bibinfo {author} {\bibfnamefont {L.}~\bibnamefont {Lavoura}}, \
  and\ \bibinfo {author} {\bibfnamefont {B.}~\bibnamefont {Radovcic}},\ }\href
  {\doibase 10.1016/j.physletb.2009.03.016} {\bibfield  {journal} {\bibinfo
  {journal} {Phys. Lett.}\ }\textbf {\bibinfo {volume} {B674}},\ \bibinfo
  {pages} {117} (\bibinfo {year} {2009})},\ \Eprint
  {http://arxiv.org/abs/0902.2325} {arXiv:0902.2325 [hep-ph]} \BibitemShut
  {NoStop}%
%%CITATION = ARXIV:0902.2325;%%
\bibitem [{\citenamefont {Wang}\ \emph {et~al.}(2006)\citenamefont {Wang},
  \citenamefont {Wang},\ and\ \citenamefont {Yang}}]{Wang:2006jy}%
  \BibitemOpen
  \bibfield  {author} {\bibinfo {author} {\bibfnamefont {F.}~\bibnamefont
  {Wang}}, \bibinfo {author} {\bibfnamefont {W.}~\bibnamefont {Wang}}, \ and\
  \bibinfo {author} {\bibfnamefont {J.~M.}\ \bibnamefont {Yang}},\ }\href
  {\doibase 10.1209/epl/i2006-10293-3} {\bibfield  {journal} {\bibinfo
  {journal} {Europhys. Lett.}\ }\textbf {\bibinfo {volume} {76}},\ \bibinfo
  {pages} {388} (\bibinfo {year} {2006})},\ \Eprint
  {http://arxiv.org/abs/hep-ph/0601018} {arXiv:hep-ph/0601018 [hep-ph]}
  \BibitemShut {NoStop}%
%%CITATION = HEP-PH/0601018;%%
\bibitem [{\citenamefont {Davidson}\ and\ \citenamefont
  {Logan}(2009)}]{Davidson:2009ha}%
  \BibitemOpen
  \bibfield  {author} {\bibinfo {author} {\bibfnamefont {S.~M.}\ \bibnamefont
  {Davidson}}\ and\ \bibinfo {author} {\bibfnamefont {H.~E.}\ \bibnamefont
  {Logan}},\ }\href {\doibase 10.1103/PhysRevD.80.095008} {\bibfield  {journal}
  {\bibinfo  {journal} {Phys. Rev.}\ }\textbf {\bibinfo {volume} {D80}},\
  \bibinfo {pages} {095008} (\bibinfo {year} {2009})},\ \Eprint
  {http://arxiv.org/abs/0906.3335} {arXiv:0906.3335 [hep-ph]} \BibitemShut
  {NoStop}%
%%CITATION = ARXIV:0906.3335;%%
\bibitem [{\citenamefont {Ma}\ and\ \citenamefont {Popov}(2017)}]{Ma:2016mwh}%
  \BibitemOpen
  \bibfield  {author} {\bibinfo {author} {\bibfnamefont {E.}~\bibnamefont
  {Ma}}\ and\ \bibinfo {author} {\bibfnamefont {O.}~\bibnamefont {Popov}},\
  }\href {\doibase 10.1016/j.physletb.2016.11.027} {\bibfield  {journal}
  {\bibinfo  {journal} {Phys. Lett.}\ }\textbf {\bibinfo {volume} {B764}},\
  \bibinfo {pages} {142} (\bibinfo {year} {2017})},\ \Eprint
  {http://arxiv.org/abs/1609.02538} {arXiv:1609.02538 [hep-ph]} \BibitemShut
  {NoStop}%
%%CITATION = ARXIV:1609.02538;%%
\bibitem [{\citenamefont {Baek}\ and\ \citenamefont
  {Nomura}(2017)}]{Baek:2016wml}%
  \BibitemOpen
  \bibfield  {author} {\bibinfo {author} {\bibfnamefont {S.}~\bibnamefont
  {Baek}}\ and\ \bibinfo {author} {\bibfnamefont {T.}~\bibnamefont {Nomura}},\
  }\href {\doibase 10.1007/JHEP03(2017)059} {\bibfield  {journal} {\bibinfo
  {journal} {JHEP}\ }\textbf {\bibinfo {volume} {03}},\ \bibinfo {pages} {059}
  (\bibinfo {year} {2017})},\ \Eprint {http://arxiv.org/abs/1611.09145}
  {arXiv:1611.09145 [hep-ph]} \BibitemShut {NoStop}%
%%CITATION = ARXIV:1611.09145;%%
\bibitem [{\citenamefont {Baek}\ \emph {et~al.}(2018)\citenamefont {Baek},
  \citenamefont {Das},\ and\ \citenamefont {Nomura}}]{Baek:2018wuo}%
  \BibitemOpen
  \bibfield  {author} {\bibinfo {author} {\bibfnamefont {S.}~\bibnamefont
  {Baek}}, \bibinfo {author} {\bibfnamefont {A.}~\bibnamefont {Das}}, \ and\
  \bibinfo {author} {\bibfnamefont {T.}~\bibnamefont {Nomura}},\ }\href
  {\doibase 10.1007/JHEP05(2018)205} {\bibfield  {journal} {\bibinfo  {journal}
  {JHEP}\ }\textbf {\bibinfo {volume} {05}},\ \bibinfo {pages} {205} (\bibinfo
  {year} {2018})},\ \Eprint {http://arxiv.org/abs/1802.08615} {arXiv:1802.08615
  [hep-ph]} \BibitemShut {NoStop}%
%%CITATION = ARXIV:1802.08615;%%
\bibitem [{\citenamefont {Kim}(1979)}]{Kim:1979if}%
  \BibitemOpen
  \bibfield  {author} {\bibinfo {author} {\bibfnamefont {J.~E.}\ \bibnamefont
  {Kim}},\ }\href {\doibase 10.1103/PhysRevLett.43.103} {\bibfield  {journal}
  {\bibinfo  {journal} {Phys. Rev. Lett.}\ }\textbf {\bibinfo {volume} {43}},\
  \bibinfo {pages} {103} (\bibinfo {year} {1979})}\BibitemShut {NoStop}%
%%CITATION = PRLTA,43,103;%%
\bibitem [{\citenamefont {Shifman}\ \emph {et~al.}(1980)\citenamefont
  {Shifman}, \citenamefont {Vainshtein},\ and\ \citenamefont
  {Zakharov}}]{Shifman:1979if}%
  \BibitemOpen
  \bibfield  {author} {\bibinfo {author} {\bibfnamefont {M.~A.}\ \bibnamefont
  {Shifman}}, \bibinfo {author} {\bibfnamefont {A.~I.}\ \bibnamefont
  {Vainshtein}}, \ and\ \bibinfo {author} {\bibfnamefont {V.~I.}\ \bibnamefont
  {Zakharov}},\ }\href {\doibase 10.1016/0550-3213(80)90209-6} {\bibfield
  {journal} {\bibinfo  {journal} {Nucl. Phys.}\ }\textbf {\bibinfo {volume}
  {B166}},\ \bibinfo {pages} {493} (\bibinfo {year} {1980})}\BibitemShut
  {NoStop}%
%%CITATION = NUPHA,B166,493;%%
\bibitem [{\citenamefont {Dine}\ \emph {et~al.}(1981)\citenamefont {Dine},
  \citenamefont {Fischler},\ and\ \citenamefont {Srednicki}}]{Dine:1981rt}%
  \BibitemOpen
  \bibfield  {author} {\bibinfo {author} {\bibfnamefont {M.}~\bibnamefont
  {Dine}}, \bibinfo {author} {\bibfnamefont {W.}~\bibnamefont {Fischler}}, \
  and\ \bibinfo {author} {\bibfnamefont {M.}~\bibnamefont {Srednicki}},\ }\href
  {\doibase 10.1016/0370-2693(81)90590-6} {\bibfield  {journal} {\bibinfo
  {journal} {Phys. Lett.}\ }\textbf {\bibinfo {volume} {104B}},\ \bibinfo
  {pages} {199} (\bibinfo {year} {1981})}\BibitemShut {NoStop}%
%%CITATION = PHLTA,104B,199;%%
\bibitem [{\citenamefont {Zhitnitsky}(1980)}]{Zhitnitsky:1980tq}%
  \BibitemOpen
  \bibfield  {author} {\bibinfo {author} {\bibfnamefont {A.~R.}\ \bibnamefont
  {Zhitnitsky}},\ }\href@noop {} {\bibfield  {journal} {\bibinfo  {journal}
  {Sov. J. Nucl. Phys.}\ }\textbf {\bibinfo {volume} {31}},\ \bibinfo {pages}
  {260} (\bibinfo {year} {1980})},\ \bibinfo {note} {[Yad.
  Fiz.31,497(1980)]}\BibitemShut {NoStop}%
%%CITATION = SJNCA,31,260;%%
\bibitem [{\citenamefont {Berezhiani}\ and\ \citenamefont
  {Khlopov}(1991)}]{Berezhiani:1989fp}%
  \BibitemOpen
  \bibfield  {author} {\bibinfo {author} {\bibfnamefont {Z.~G.}\ \bibnamefont
  {Berezhiani}}\ and\ \bibinfo {author} {\bibfnamefont {M.~{\relax Yu}.}\
  \bibnamefont {Khlopov}},\ }\href {\doibase 10.1007/BF01570798} {\bibfield
  {journal} {\bibinfo  {journal} {Z. Phys.}\ }\textbf {\bibinfo {volume}
  {C49}},\ \bibinfo {pages} {73} (\bibinfo {year} {1991})}\BibitemShut
  {NoStop}%
%%CITATION = ZEPYA,C49,73;%%
\bibitem [{\citenamefont {Gu}\ and\ \citenamefont {He}(2006)}]{Gu:2006dc}%
  \BibitemOpen
  \bibfield  {author} {\bibinfo {author} {\bibfnamefont {P.-H.}\ \bibnamefont
  {Gu}}\ and\ \bibinfo {author} {\bibfnamefont {H.-J.}\ \bibnamefont {He}},\
  }\href {\doibase 10.1088/1475-7516/2006/12/010} {\bibfield  {journal}
  {\bibinfo  {journal} {JCAP}\ }\textbf {\bibinfo {volume} {0612}},\ \bibinfo
  {pages} {010} (\bibinfo {year} {2006})},\ \Eprint
  {http://arxiv.org/abs/hep-ph/0610275} {arXiv:hep-ph/0610275 [hep-ph]}
  \BibitemShut {NoStop}%
%%CITATION = HEP-PH/0610275;%%
\bibitem [{\citenamefont {Chen}\ and\ \citenamefont
  {Tsai}(2013)}]{Chen:2012baa}%
  \BibitemOpen
  \bibfield  {author} {\bibinfo {author} {\bibfnamefont {C.-S.}\ \bibnamefont
  {Chen}}\ and\ \bibinfo {author} {\bibfnamefont {L.-H.}\ \bibnamefont
  {Tsai}},\ }\href {\doibase 10.1103/PhysRevD.88.055015} {\bibfield  {journal}
  {\bibinfo  {journal} {Phys. Rev.}\ }\textbf {\bibinfo {volume} {D88}},\
  \bibinfo {pages} {055015} (\bibinfo {year} {2013})},\ \Eprint
  {http://arxiv.org/abs/1210.6264} {arXiv:1210.6264 [hep-ph]} \BibitemShut
  {NoStop}%
%%CITATION = ARXIV:1210.6264;%%
\bibitem [{\citenamefont {Dasgupta}\ \emph {et~al.}(2014)\citenamefont
  {Dasgupta}, \citenamefont {Ma},\ and\ \citenamefont
  {Tsumura}}]{Dasgupta:2013cwa}%
  \BibitemOpen
  \bibfield  {author} {\bibinfo {author} {\bibfnamefont {B.}~\bibnamefont
  {Dasgupta}}, \bibinfo {author} {\bibfnamefont {E.}~\bibnamefont {Ma}}, \ and\
  \bibinfo {author} {\bibfnamefont {K.}~\bibnamefont {Tsumura}},\ }\href
  {\doibase 10.1103/PhysRevD.89.041702} {\bibfield  {journal} {\bibinfo
  {journal} {Phys. Rev.}\ }\textbf {\bibinfo {volume} {D89}},\ \bibinfo {pages}
  {041702} (\bibinfo {year} {2014})},\ \Eprint {http://arxiv.org/abs/1308.4138}
  {arXiv:1308.4138 [hep-ph]} \BibitemShut {NoStop}%
%%CITATION = ARXIV:1308.4138;%%
\bibitem [{\citenamefont {Ma}(2015)}]{Ma:2014yka}%
  \BibitemOpen
  \bibfield  {author} {\bibinfo {author} {\bibfnamefont {E.}~\bibnamefont
  {Ma}},\ }\href {\doibase 10.1016/j.physletb.2014.12.045} {\bibfield
  {journal} {\bibinfo  {journal} {Phys. Lett.}\ }\textbf {\bibinfo {volume}
  {B741}},\ \bibinfo {pages} {202} (\bibinfo {year} {2015})},\ \Eprint
  {http://arxiv.org/abs/1411.6679} {arXiv:1411.6679 [hep-ph]} \BibitemShut
  {NoStop}%
%%CITATION = ARXIV:1411.6679;%%
\bibitem [{\citenamefont {Bertolini}\ \emph {et~al.}(2015)\citenamefont
  {Bertolini}, \citenamefont {Di~Luzio}, \citenamefont {Kole\v{s}ov\'{a}},\
  and\ \citenamefont {Malinsk\'{y}}}]{Bertolini:2014aia}%
  \BibitemOpen
  \bibfield  {author} {\bibinfo {author} {\bibfnamefont {S.}~\bibnamefont
  {Bertolini}}, \bibinfo {author} {\bibfnamefont {L.}~\bibnamefont {Di~Luzio}},
  \bibinfo {author} {\bibfnamefont {H.}~\bibnamefont {Kole\v{s}ov\'{a}}}, \
  and\ \bibinfo {author} {\bibfnamefont {M.}~\bibnamefont {Malinsk\'{y}}},\
  }\href {\doibase 10.1103/PhysRevD.91.055014} {\bibfield  {journal} {\bibinfo
  {journal} {Phys. Rev.}\ }\textbf {\bibinfo {volume} {D91}},\ \bibinfo {pages}
  {055014} (\bibinfo {year} {2015})},\ \Eprint {http://arxiv.org/abs/1412.7105}
  {arXiv:1412.7105 [hep-ph]} \BibitemShut {NoStop}%
%%CITATION = ARXIV:1412.7105;%%
\bibitem [{\citenamefont {Ahn}\ and\ \citenamefont {Chun}(2016)}]{Ahn:2015pia}%
  \BibitemOpen
  \bibfield  {author} {\bibinfo {author} {\bibfnamefont {Y.~H.}\ \bibnamefont
  {Ahn}}\ and\ \bibinfo {author} {\bibfnamefont {E.~J.}\ \bibnamefont {Chun}},\
  }\href {\doibase 10.1016/j.physletb.2015.11.067} {\bibfield  {journal}
  {\bibinfo  {journal} {Phys. Lett.}\ }\textbf {\bibinfo {volume} {B752}},\
  \bibinfo {pages} {333} (\bibinfo {year} {2016})},\ \Eprint
  {http://arxiv.org/abs/1510.01015} {arXiv:1510.01015 [hep-ph]} \BibitemShut
  {NoStop}%
%%CITATION = ARXIV:1510.01015;%%
\bibitem [{\citenamefont {Gu}(2016)}]{Gu:2016hxh}%
  \BibitemOpen
  \bibfield  {author} {\bibinfo {author} {\bibfnamefont {P.-H.}\ \bibnamefont
  {Gu}},\ }\href {\doibase 10.1088/1475-7516/2016/07/004} {\bibfield  {journal}
  {\bibinfo  {journal} {JCAP}\ }\textbf {\bibinfo {volume} {1607}},\ \bibinfo
  {pages} {004} (\bibinfo {year} {2016})},\ \Eprint
  {http://arxiv.org/abs/1603.05070} {arXiv:1603.05070 [hep-ph]} \BibitemShut
  {NoStop}%
%%CITATION = ARXIV:1603.05070;%%
\bibitem [{\citenamefont {Ma}\ \emph {et~al.}(2018)\citenamefont {Ma},
  \citenamefont {Restrepo},\ and\ \citenamefont {Zapata}}]{Ma:2017zyb}%
  \BibitemOpen
  \bibfield  {author} {\bibinfo {author} {\bibfnamefont {E.}~\bibnamefont
  {Ma}}, \bibinfo {author} {\bibfnamefont {D.}~\bibnamefont {Restrepo}}, \ and\
  \bibinfo {author} {\bibfnamefont {{\'{O}}.}~\bibnamefont {Zapata}},\ }\href
  {\doibase 10.1142/S0217732318500244} {\bibfield  {journal} {\bibinfo
  {journal} {Mod. Phys. Lett.}\ }\textbf {\bibinfo {volume} {A33}},\ \bibinfo
  {pages} {1850024} (\bibinfo {year} {2018})},\ \Eprint
  {http://arxiv.org/abs/1706.08240} {arXiv:1706.08240 [hep-ph]} \BibitemShut
  {NoStop}%
%%CITATION = ARXIV:1706.08240;%%
\bibitem [{\citenamefont {Suematsu}(2018)}]{Suematsu:2017kcu}%
  \BibitemOpen
  \bibfield  {author} {\bibinfo {author} {\bibfnamefont {D.}~\bibnamefont
  {Suematsu}},\ }\href {\doibase 10.1140/epjc/s10052-018-5519-4} {\bibfield
  {journal} {\bibinfo  {journal} {Eur. Phys. J.}\ }\textbf {\bibinfo {volume}
  {C78}},\ \bibinfo {pages} {33} (\bibinfo {year} {2018})},\ \Eprint
  {http://arxiv.org/abs/1709.02886} {arXiv:1709.02886 [hep-ph]} \BibitemShut
  {NoStop}%
%%CITATION = ARXIV:1709.02886;%%
\bibitem [{\citenamefont {Ahn}(2018)}]{Ahn:2018cau}%
  \BibitemOpen
  \bibfield  {author} {\bibinfo {author} {\bibfnamefont {Y.~H.}\ \bibnamefont
  {Ahn}},\ }\href {\doibase 10.1103/PhysRevD.98.035047} {\bibfield  {journal}
  {\bibinfo  {journal} {Phys. Rev.}\ }\textbf {\bibinfo {volume} {D98}},\
  \bibinfo {pages} {035047} (\bibinfo {year} {2018})},\ \Eprint
  {http://arxiv.org/abs/1804.06988} {arXiv:1804.06988 [hep-ph]} \BibitemShut
  {NoStop}%
%%CITATION = ARXIV:1804.06988;%%
\bibitem [{\citenamefont {Reig}\ and\ \citenamefont
  {Srivastava}(2019)}]{Reig:2018yfd}%
  \BibitemOpen
  \bibfield  {author} {\bibinfo {author} {\bibfnamefont {M.}~\bibnamefont
  {Reig}}\ and\ \bibinfo {author} {\bibfnamefont {R.}~\bibnamefont
  {Srivastava}},\ }\href {\doibase 10.1016/j.physletb.2019.01.008} {\bibfield
  {journal} {\bibinfo  {journal} {Phys. Lett.}\ }\textbf {\bibinfo {volume}
  {B790}},\ \bibinfo {pages} {134} (\bibinfo {year} {2019})},\ \Eprint
  {http://arxiv.org/abs/1809.02093} {arXiv:1809.02093 [hep-ph]} \BibitemShut
  {NoStop}%
%%CITATION = ARXIV:1809.02093;%%
\bibitem [{\citenamefont {Carvajal}\ and\ \citenamefont
  {Zapata}(2019)}]{Carvajal:2018ohk}%
  \BibitemOpen
  \bibfield  {author} {\bibinfo {author} {\bibfnamefont {C.~D.~R.}\
  \bibnamefont {Carvajal}}\ and\ \bibinfo {author} {\bibfnamefont
  {{\'{O}}.}~\bibnamefont {Zapata}},\ }\href {\doibase
  10.1103/PhysRevD.99.075009} {\bibfield  {journal} {\bibinfo  {journal} {Phys.
  Rev.}\ }\textbf {\bibinfo {volume} {D99}},\ \bibinfo {pages} {075009}
  (\bibinfo {year} {2019})},\ \Eprint {http://arxiv.org/abs/1812.06364}
  {arXiv:1812.06364 [hep-ph]} \BibitemShut {NoStop}%
%%CITATION = ARXIV:1812.06364;%%
\bibitem [{\citenamefont {Kannike}(2012)}]{Kannike:2012pe}%
  \BibitemOpen
  \bibfield  {author} {\bibinfo {author} {\bibfnamefont {K.}~\bibnamefont
  {Kannike}},\ }\href {\doibase 10.1140/epjc/s10052-012-2093-z} {\bibfield
  {journal} {\bibinfo  {journal} {Eur. Phys. J.}\ }\textbf {\bibinfo {volume}
  {C72}},\ \bibinfo {pages} {2093} (\bibinfo {year} {2012})},\ \Eprint
  {http://arxiv.org/abs/1205.3781} {arXiv:1205.3781 [hep-ph]} \BibitemShut
  {NoStop}%
%%CITATION = ARXIV:1205.3781;%%
\bibitem [{\citenamefont {Foot}\ \emph {et~al.}(2014)\citenamefont {Foot},
  \citenamefont {Kobakhidze}, \citenamefont {McDonald},\ and\ \citenamefont
  {Volkas}}]{Foot:2013hna}%
  \BibitemOpen
  \bibfield  {author} {\bibinfo {author} {\bibfnamefont {R.}~\bibnamefont
  {Foot}}, \bibinfo {author} {\bibfnamefont {A.}~\bibnamefont {Kobakhidze}},
  \bibinfo {author} {\bibfnamefont {K.~L.}\ \bibnamefont {McDonald}}, \ and\
  \bibinfo {author} {\bibfnamefont {R.~R.}\ \bibnamefont {Volkas}},\ }\href
  {\doibase 10.1103/PhysRevD.89.115018} {\bibfield  {journal} {\bibinfo
  {journal} {Phys. Rev.}\ }\textbf {\bibinfo {volume} {D89}},\ \bibinfo {pages}
  {115018} (\bibinfo {year} {2014})},\ \Eprint {http://arxiv.org/abs/1310.0223}
  {arXiv:1310.0223 [hep-ph]} \BibitemShut {NoStop}%
%%CITATION = ARXIV:1310.0223;%%
\bibitem [{\citenamefont {Srednicki}(1985)}]{Srednicki:1985xd}%
  \BibitemOpen
  \bibfield  {author} {\bibinfo {author} {\bibfnamefont {M.}~\bibnamefont
  {Srednicki}},\ }\href {\doibase 10.1016/0550-3213(85)90054-9} {\bibfield
  {journal} {\bibinfo  {journal} {Nucl. Phys.}\ }\textbf {\bibinfo {volume}
  {B260}},\ \bibinfo {pages} {689} (\bibinfo {year} {1985})}\BibitemShut
  {NoStop}%
%%CITATION = NUPHA,B260,689;%%
\bibitem [{\citenamefont {Peinado}\ \emph {et~al.}(2019)\citenamefont
  {Peinado}, \citenamefont {Reig}, \citenamefont {Srivastava},\ and\
  \citenamefont {Valle}}]{Peinado:2019mrn}%
  \BibitemOpen
  \bibfield  {author} {\bibinfo {author} {\bibfnamefont {E.}~\bibnamefont
  {Peinado}}, \bibinfo {author} {\bibfnamefont {M.}~\bibnamefont {Reig}},
  \bibinfo {author} {\bibfnamefont {R.}~\bibnamefont {Srivastava}}, \ and\
  \bibinfo {author} {\bibfnamefont {J.~W.~F.}\ \bibnamefont {Valle}},\
  }\href@noop {} {\  (\bibinfo {year} {2019})},\ \Eprint
  {http://arxiv.org/abs/1910.02961} {arXiv:1910.02961 [hep-ph]} \BibitemShut
  {NoStop}%
%%CITATION = ARXIV:1910.02961;%%
\bibitem [{\citenamefont {Grilli~di Cortona}\ \emph {et~al.}(2016)\citenamefont
  {Grilli~di Cortona}, \citenamefont {Hardy}, \citenamefont {Pardo~Vega},\ and\
  \citenamefont {Villadoro}}]{diCortona:2015ldu}%
  \BibitemOpen
  \bibfield  {author} {\bibinfo {author} {\bibfnamefont {G.}~\bibnamefont
  {Grilli~di Cortona}}, \bibinfo {author} {\bibfnamefont {E.}~\bibnamefont
  {Hardy}}, \bibinfo {author} {\bibfnamefont {J.}~\bibnamefont {Pardo~Vega}}, \
  and\ \bibinfo {author} {\bibfnamefont {G.}~\bibnamefont {Villadoro}},\ }\href
  {\doibase 10.1007/JHEP01(2016)034} {\bibfield  {journal} {\bibinfo  {journal}
  {JHEP}\ }\textbf {\bibinfo {volume} {01}},\ \bibinfo {pages} {034} (\bibinfo
  {year} {2016})},\ \Eprint {http://arxiv.org/abs/1511.02867} {arXiv:1511.02867
  [hep-ph]} \BibitemShut {NoStop}%
%%CITATION = ARXIV:1511.02867;%%
\bibitem [{\citenamefont {Tanabashi}\ \emph {et~al.}(2018)\citenamefont
  {Tanabashi} \emph {et~al.}}]{Tanabashi:2018oca}%
  \BibitemOpen
  \bibfield  {author} {\bibinfo {author} {\bibfnamefont {M.}~\bibnamefont
  {Tanabashi}} \emph {et~al.} (\bibinfo {collaboration} {Particle Data
  Group}),\ }\href {\doibase 10.1103/PhysRevD.98.030001} {\bibfield  {journal}
  {\bibinfo  {journal} {Phys. Rev.}\ }\textbf {\bibinfo {volume} {D98}},\
  \bibinfo {pages} {030001} (\bibinfo {year} {2018})}\BibitemShut {NoStop}%
%%CITATION = PHRVA,D98,030001;%%
\bibitem [{\citenamefont {Huang}\ and\ \citenamefont
  {Nath}(2018)}]{Huang:2018cwo}%
  \BibitemOpen
  \bibfield  {author} {\bibinfo {author} {\bibfnamefont {G.-Y.}\ \bibnamefont
  {Huang}}\ and\ \bibinfo {author} {\bibfnamefont {N.}~\bibnamefont {Nath}},\
  }\href {\doibase 10.1140/epjc/s10052-018-6391-y} {\bibfield  {journal}
  {\bibinfo  {journal} {Eur. Phys. J.}\ }\textbf {\bibinfo {volume} {C78}},\
  \bibinfo {pages} {922} (\bibinfo {year} {2018})},\ \Eprint
  {http://arxiv.org/abs/1809.01111} {arXiv:1809.01111 [hep-ph]} \BibitemShut
  {NoStop}%
%%CITATION = ARXIV:1809.01111;%%
\end{thebibliography}%

\end{document}